\documentclass[11pt,leqno]{article}

\usepackage[top=3cm, bottom=3cm, left=4cm, right=4cm ]{geometry}

\usepackage{amsmath,amssymb,amsthm}
\usepackage{enumitem}

\newcommand{\mfT}{\mathfrak{T}}
\newcommand{\mft}{\mathfrak{t}}
\newcommand{\mfG}{\mathfrak{G}}
\newcommand{\mfA}{\mathfrak{A}}
\newcommand{\mfg}{\mathfrak{g}}

\begin{document}

\title{Radiation and Boundary Conditions in the Theory of Gravitation\footnote{This is an author-created version of an article presented by L. Infeld on April 12, 1958 to and published in Bulletin de l'Acad\'emie Polonaise des Sciences, S\'erie des sci. math., astr. et phys. Vol. {\bf VI}, No. 6 (1958) 407-412.}}
\author{by\medskip\\Andrzej Trautman}
%\date{April 12, 1958}
\date{\empty}

\maketitle

   The aim of this paper is to discuss the connection between the problem of gravitational radiation and the boundary conditions at infinity. 
We shall deal with the concept of energy and momentum in Einstein's 
general relativity and propose a prescription for computing the total 
radiated energy. A connection between our radiation conditions and the 
definitions of gravitational radiation by Pirani and Lichnerowicz is shown 
in section 5.

\paragraph{1.} In physics we are ordinarily interested in conservation laws which 
have an integral character. A classical conserved quantity is a functional 
 $f[\sigma]$  depending on a space-like hypersurface $\sigma$. A conservation law is 
a statement that, by virtue of the equations of motion, $f$, in fact, does 
not depend on $\sigma$. As is known, in general relativity the energy-momentum tensor of matter $\mfT_{\mu\nu}$ does not by itself lead to an integral conservation law. However, if we introduce an energy-momentum pseudotensor of the gravitational field $\mft^\nu_\mu=(\delta^\nu_\mu\mfG+g^{\rho\sigma}{}_{,\mu}\partial\mfG/\partial g^{\rho\sigma}{}_{,\nu})/2\varkappa$, then the sum 
$\mfT^\nu_\mu+\mft^\nu_\mu$ is divergenceless by virtue of Einstein's equations\footnote{We shall use the notations of the preceding paper; $g_{\mu\nu}$  will denote the metric 
tensor of the Riemannian space-time $V_{4}$. Gothic letters denote tensor densities and 
also ``pseudoquantities'' such as the superpotentials.}. Einstein's tensor density $\mfG^\nu{}_\mu=\sqrt{-g}(R^\nu_\mu-\frac{1}{2}\delta^\nu_\mu R)$  can namely be written in the form
\begin{equation}
\mfG_\mu{}^\nu\equiv\varkappa(\mft_\mu{}^\nu+\mfA_\mu{}^{\lambda\nu}{}_{,\lambda}),
\label{G}
\end{equation}
where the ``superpotentials'' $\mfA_\mu{}^{\nu\lambda}$  are given in \cite{gold}
\begin{equation}
2\varkappa\mfA_\mu{}^{\nu\lambda}\equiv \sqrt{-g}g^{\sigma[\rho}\delta^\nu_\mu g^{\lambda]\tau}g_{\rho\sigma,\tau}\equiv-2\varkappa\mfA_\mu{}^{\lambda\nu}.
\label{A}
\end{equation}

If the Einstein equations
\begin{equation}
G_{\mu\nu}=-\varkappa T_{\mu\nu}
\label{G=T}
\end{equation}
are satisfied, then Eqs. \eqref{G} and \eqref{A} imply
\begin{equation}
\mfT_\mu{}^\nu+\mft_\mu{}^\nu=\mfA_\mu{}^{\nu\lambda}{}_{,\lambda},\quad\quad\text{thus}\quad\quad (\mfT_\mu{}^\nu+\mft_\mu{}^\nu)_{,\nu}=0.
\label{TtA}
\end{equation}

The functions $\mft_\mu{}^\nu$  are not components of a tensor density (equivalence 
principle) and many physicists (e. g., Schr\"odinger \cite{schr}) have raised doubts 
as to their physical meaning. Einstein \cite{ein} and F. Klein \cite{klein} formulated some conditions which enable us to consider the integrals
\begin{equation}
P_\mu[\sigma]=\int_\sigma(\mfT_\mu{}^\nu+\mft_\mu{}^\nu)\,dS_\nu=\oint_S\mfA_\mu{}^{\nu\lambda}\,dS_{\nu\lambda}
\label{Pmu}
\end{equation}
as representing the total energy and momentum of the system: matter 
and gravitational field. These conditions can be summarized as follows. 
Let us take an isolated system of masses ($T_{\mu\nu}=0$  outside a bounded 
3-region) and assume the existence of co-ordinates such that \cite{lich-1}
\begin{align}
g_{\mu\nu}&=\eta_{\mu\nu}+O(r^{-1}),& g_{\mu\nu,\rho}&=O(r^{-2}),
\label{g-eta}
\end{align}
where $r$ denotes the distance measured along geodesics from a fixed point 
on a space-like $\sigma$. Eqs. \eqref{g-eta} have a double meaning: they constitute a system of boundary conditions, and they distinguish a set of co-ordinate 
systems (``Galilean at infinity'').

     Using \eqref{TtA} it can be easily proved that: $(i)$  $P_\mu[\sigma]$, calculated from \eqref{Pmu} in a co-ordinate system satisfying \eqref{g-eta}, is always finite and does not depend on $\sigma$; $(ii)$ $P_\mu$  does not depend on co-ordinate changes which do 
not alter \eqref{g-eta} and reduce to an identity for $r\to\infty$; $(iii)$ $P_\mu$  is a vector 
with respect to linear orthogonal transformations. The proof is based 
on the vanishing of the integral
\begin{equation}
p_\mu=\int_\Sigma (\mfT_\mu{}^\nu+\mft_\mu{}^\nu)\,dS_\nu
\label{pmu}
\end{equation}
taken over a {\em time-like} ``cylindrical'' hypersurface $\Sigma$  at spatial infinity (note that $S$  appearing in \eqref{Pmu}  is the intersection of $\Sigma$  and $\sigma$). The vanishing of these integrals is ensured by \eqref{g-eta} ($\mft_\mu{}^\nu$  is quadratic in $g_{\mu\nu,\rho}$) and our assumption on $T_{\mu\nu}$. The integral \eqref{pmu} can eventually be identified 
with the total energy and momentum radiated through $\Sigma$, and Lichnerowicz's boundary conditions \eqref{g-eta} automatically exclude the existence of 
any radiation.

\paragraph{2.} Comparison with electrodynamics suggests that radiation fields 
in general relativity should be characterized by $g_{\mu\nu,\rho}\sim 1/r$, rather than by $g_{\mu\nu,\rho}\sim 1/r^2$. However, if the integrals \eqref{pmu} do not vanish, the proof of the Einstein-Klein theorem is no longer valid and doubts as to the
meaning of \eqref{Pmu} arise anew. We propose to generalize the boundary conditions \eqref{g-eta} in such a way as to include radiation fields. We expect that 
these conditions will ensure the finiteness of $P_\mu$  and that $P_\mu$  will not 
change with co-ordinate transformations which reduce to an identity 
for $r\to\infty$  and preserve the {\em form} of the boundary conditions. The dependence of $P_\mu$  on $\sigma$  will now correspond to the diminishing of total energy due to radiation.

Fock \cite{fock} proposes to normalize the co-ordinate systems by means 
of de Donder's relation
\begin{equation}
\mfg^{\mu\nu}{}_{,\nu}=0
\label{g^}
\end{equation}
and imposes on $g_{\mu\nu}$  the radiation condition of Sommerfeld. We find this 
formulation somewhat stringent. In particular, we see no reason for 
restricting ourselves to harmonic co-ordinates only. There is no convincing argument for writing the Schwarzschild line element in harmonic 
co-ordinates instead of, say, in isotropic ones.

We generalize the conditions of Fock along the lines presented in 
the preceding paper. First, introduce a null vector field $k_\nu$, defined as 
follows. Let $n^\nu$  be a unit space-like vector lying in $\sigma$, perpendicular to 
the ``sphere'' $r={\rm const.}$, and pointing outside it. We put $k^\nu=n^\nu+t^\nu$, where $t$  denotes a unit time-like vector normal to $\sigma$, such that $t^0> 0$.

    Now, we formulate the following boundary conditions to be imposed 
on gravitational fields due to isolated systems of matter: {\em there exist 
co-ordinate systems and functions $h_{\mu\nu}=O(r^{-1})$  such that}
\begin{gather}
\begin{aligned}
g_{\mu\nu}&=\eta_{\mu\nu}+O(r^{-1}),\quad\quad& g_{\mu\nu,\rho}&=h_{\mu\nu}k_\rho+O(r^{-2}),\label{h-hk}
\end{aligned}\\
(h_{\mu\nu}-\frac{1}{2}\eta_{\mu\nu}\eta^{\rho\sigma}h_{\rho\sigma})k^\nu=O(r^{-2}).\label{hhk}
\end{gather}

These conditions correspond to Sommerfeld's ``Ausstrahlungsbe\-din\-gung''; we obtain the ``Einstrahlungsbedingung''
assuming $n^\nu$ to be a normal pointing inward the sphere $r={\rm
  const.}$ Relations \eqref{h-hk}, \eqref{hhk} are {\em weaker} than \eqref{g-eta}; this means that every field fulfilling \eqref{g-eta} satisfies also 
our conditions \eqref{h-hk}, \eqref{hhk}. The class of co-ordinate systems distinguished by \eqref{h-hk} and \eqref{hhk} is larger than that defined by \eqref{g-eta}. Eq. \eqref{hhk} restricts the co-ordinate systems to those which are asymptotically harmonic; however, it must be noted that isotropic co-ordinates used in the Schwarzschild $V_4$ are asymptotically harmonic in our meaning.

Strictly speaking the correctness of conditions \eqref{h-hk} and \eqref{hhk} might 
be inferred only if it were possible to show that Einstein's equations 
with bounded sources have always exactly one solution satisfying \eqref{h-hk}, 
\eqref{hhk}. But it is not an easy task to prove this theorem.

\paragraph{3.} We shall now present some consequences of \eqref{h-hk} and \eqref{hhk}. First of all, we must examine the convergence of energy integrals \eqref{Pmu}. The superpotentials are linear in $g_{\mu\nu,\rho}$  and thus go as $1/r$; we must therefore show that the terms behaving as $1/r$ cancel out in the surface integral \eqref{Pmu}. Indeed, the surface element $dS_{\lambda\nu}$ is proportional to $n_{[\lambda}t_{\nu]}=n_{[\lambda}k_{\nu]}$, and the terms in question in \eqref{Pmu} can be written as $\eta^{\sigma[\rho}\delta^\nu_\mu\eta^{\lambda]\tau}h_{\rho\sigma}k_\tau k_{[\nu}n_{\lambda]}$. Taking into account \eqref{hhk}, we verify that this expression does vanish.

Let us take a co-ordinate transformation
\begin{equation}
x^\nu\to x^{\prime\nu}=x^\nu+a^\nu(x)
\label{xx}
\end{equation}
fulfilling
\begin{align}
a^\nu&=o(r),& a_{\nu,\mu}&=b_\nu k_\mu+O(r^{-2}) %czy nie powinno byc a^\nu&=O(r) ?
\label{aa}
\end{align}
where
\begin{align*}
a_\nu&=\eta_{\nu\mu}a^\mu,& b_\nu=O(r^{-1}),
\end{align*}
and
\begin{align}
a_{\nu,\mu\rho}&=b_{\nu,\mu}k_\rho+O(r^{-2}),& b_{\nu,\rho}&=O(r^{-1}).
\label{ab}
\end{align}

From (13) follows the existence of functions $c_\nu=O(r^{-1})$   such that
\begin{equation}
b_{\nu,\mu}=c_\nu k_\mu+O(r^{-2}).
\label{b-ck}
\end{equation}

Co-ordinate transformations \eqref{xx} satisfying \eqref{aa} and \eqref{ab} preserve the form of our boundary conditions; this can be easily seen from the 
transformation formulae for $g_{\mu\nu}$  and $h_{\mu\nu}$:
\begin{equation}
\begin{aligned}
g'_{\mu\nu}(x')&\cong g_{\mu\nu}(x)+b_\mu k_\nu+b_\nu k_\mu,\\
h'_{\mu\nu}(x')&\cong h_{\mu\nu}(x)+c_\mu k_\nu+c_\nu k_\mu.
\end{aligned}
\label{gh}
\end{equation}

Computing the superpotentials in both co-ordinate systems and 
taking into account the relations \eqref{h-hk}-\eqref{gh} we obtain
\[
\mfA'_{\mu}{}^{\nu\lambda}k'_\nu n'_\lambda=\mfA_{\mu}{}^{\nu\lambda}k_\nu n_\lambda+O(r^{-3}).
\]
Therefore, the total energy and momentum $P_\mu$ are well defined by \eqref{Pmu} and the boundary conditions \eqref{h-hk}, \eqref{hhk}. It must be noted that our prescription demands that the {\em calculation} of $P_\mu$  be performed by 
means of \eqref{Pmu} using co-ordinates which satisfy Eqs. \eqref{h-hk} and \eqref{hhk}. This does not at all mean that the energy is only a property of the co-ordinate system. The vector $P_\mu[\sigma]$ constitutes a {\em global} characteristic of the field and it is only for computational purposes that we must appeal to \eqref{h-hk}, \eqref{hhk}.

\paragraph{4.} The total energy and momentum $p_\mu$ radiated between two hypersurfaces $\sigma$ and $\sigma'$ is given by \eqref{pmu}, or by
\[
p_\mu=P_\mu[\sigma]-P_\mu[\sigma']=\int_\Sigma \mft_\mu{}^\nu\,dS_\nu
\]
($T_{\mu\nu}$ vanishes on $\Sigma$). The boundary conditions enable the estimation of $p_\mu$; we have, indeed,
\begin{equation}
\mft_\mu{}^{\nu}=\tau k_\mu k^\nu+O(r^{-3}),
\label{t-kk}
\end{equation}
where
\begin{equation}
4\varkappa\tau=h^{\mu\nu}(h_{\mu\nu}-\frac{1}{2}\eta_{\mu\nu}\eta^{\rho\sigma}h_{\rho\sigma}).
\label{4ktau}
\end{equation}
$\tau$ is invariant with respect to transformation \eqref{gh} and is
{\em non-negative} by virtue of \eqref{hhk}; therefore $p_0\geq 0$. The existence of radiation is characterized by $p_\mu\neq 0$.

We could also take a more general case, including the electromagnetic field. The boundary conditions for $g_{\mu\nu}$, should be supplemented by those for $f_{\rho\sigma}$ given in the preceding paper.

We obtain in this case $\mfT_\mu{}^{\nu}+\mft_\mu{}^{\nu}=\bar{\tau}k_\mu k^\nu+O(r^{-3})$, $0\leq\bar{\tau}=O(r^{-2})$.  

\paragraph{5.} Pirani \cite{pir} and Lichnerowicz \cite{lich-2} recently proposed definitions of pure radiation fields. It may be interesting to compare their definitions with our approach. Let us admit the additional but reasonable assumption that the second derivatives of $g_{\mu\nu}$ also go to $0$  as $1/r$ and that
$g_{\mu\nu,\rho\sigma}\cong h_{\mu\nu,\sigma}k_\rho$. From $h_{\mu\nu,\sigma}k_\rho\cong h_{\mu\nu,\rho}k_\sigma$ there follows the existence of functions $i_{\mu\nu}=O(r^{-1})$  such that
\begin{align}
g_{\mu\nu,\rho\sigma}&\cong i_{\mu\nu}k_\rho k_\sigma, & (i_{\mu\nu}-\frac{1}{2}\eta_{\mu\nu}\eta^{\rho\sigma}i_{\rho\sigma})k^\nu&\cong 0.
\label{g-ikk}
\end{align}
For the curvature tensor we get
\begin{equation}
R\cong\frac{1}{2}k_{[\mu}i_{\nu][\rho}k_{\sigma]}.
\label{R}
\end{equation}
The principal part of $R_{\mu\nu\rho\sigma}$  has therefore the same form as a discontinuity of the Riemann tensor \cite{amt} and is thus of type II in the Petrov-Pirani classification \cite{pir}.

The terms proportional to $1/r$ in $R_{\mu\nu}$  cancel out by virtue of \eqref{hhk}. Conversely, $R_{\mu\nu}\cong 0$  and Eq. \eqref{g-ikk} imply $R_{\mu\nu\rho\sigma}\cong 0$ unless $k_\nu k^\nu=0$. If we take into account the electromagnetic field, Einstein's equations can be 
written in the form
\begin{align}
R_{\mu\nu}&=\rho k_\mu k_\nu +O(r^{-3}), & \rho&=O(r^{-2}.)
\label{R-rkk}
\end{align}

Moreover, it follows from \eqref{R} that
\begin{align}
k_{[\mu}R_{\nu\rho]\sigma\tau}&\cong 0, & k^\mu R_{\mu\nu\rho\sigma}&\cong 0.
\label{kR}
\end{align}
If one replaces the asymptotic equalities $\cong$ by strict ones, then Eqs. \eqref{R-rkk} and \eqref{kR} become Lichnerowicz's conditions \cite{lich-2} characterizing a pure radiation field. The definitions of Lichnerowicz and Pirani concern the idealized case of pure radiation. Actual metrics approach these radiation fields only in the limit $r\to\infty$ (wave zone).\medskip

The author is greatly indebted to Professor L. Infeld for his kind 
interest in this work. Thanks are also due to Dr. F. Pirani and W. Tulczyjew for stimulating discussions.


\begin{thebibliography}{}
\bibitem{gold} J. Goldberg, Phys. Rev., {\bf 89} (1953), 263.
\bibitem{schr} E. Schr\"odinger, Physik. Z., {\bf 19} (1918), 4.
\bibitem{ein} A. Einstein, Berl. Ber., (1918), 448.
\bibitem{klein} F. Klein, Nachr. Ges. G\"ottingen (1918), 394.
\bibitem{lich-1} A. Lichnerowicz, {\em Th\'eories relativistes de la gravitation et de l'\'ele\-ctro\-mag\-n\'e\-tis\-me}, Paris 1955.
\bibitem{fock} V. Fock, {\em Theory of space, time and gravitation} (in Russian), Moscow, 1955.
\bibitem{pir} F. Pirani, Phys. Rev. {\bf 105} (1957), 1089.
\bibitem{lich-2} A. Lichnerowicz, Comptes Rendus, {\bf 246} (1958), 893.
\bibitem{amt} A. Trautman, {\em Discontinuities of field derivatives and radiation in covariant 
theories}, Bull. Acad. Polon. Sci. Cl. III, {\bf 5} (1957), 273.
\end{thebibliography}
\end{document}